\documentclass[final,5p,times,twocolumn]{elsarticle}

\usepackage{graphics}
\usepackage{graphicx}

\usepackage{amssymb}

\journal{**}

\begin{document}

\begin{frontmatter}



\title{Novel Superconducting Phases in Copper Oxides and Iron-oxypnictides: \\NMR Studies}


 \author[label1]{Yoshio Kitaoka}
 \ead{kitaoka@mp.es.osaka-u.ac.jp}
 \author[label1]{Hidekazu Mukuda}
 \author[label1]{Sunao Shimizu}
 \author[label1]{Shin-ichiro Tabata}

 \author[label2]{Parasharam M. Shirage}
 \author[label2]{Akira Iyo}
 
\address[label1]{Graduate School of Engineering Science, Osaka University, Toyonaka, Osaka 560-8531, Japan}
\address[label2]{National Institute of Advanced Industrial Science and Technology, Tsukuba, Ibaraki, 305-8568, Japan}

\begin{abstract}

We reexamine the novel phase diagrams of antiferromagnetism (AFM) and high-$T_c$ superconductivity (HTSC) for a disorder-free CuO$_2$ plane based on an evaluation of local hole density ($p$) by site-selective Cu-NMR studies on multilayered copper oxides. Multilayered systems provide us with the opportunity to research the characteristics of the disorder-free CuO$_2$ plane. The site-selective NMR is the best and the only tool used to extract layer-dependent characteristics. Consequently, we have concluded that the uniform mixing of AFM and SC is a general property inherent to a single CuO$_2$ plane in an underdoped regime of HTSC. The $T$=0 phase diagram of AFM constructed here is in quantitative agreement with the theories in a strong correlation regime which is unchanged even with mobile holes. This~{\it Mott physics} plays a vital role for mediating the Cooper pairs to make $T_c$ of HTSC very high.
By contrast, we address from extensive NMR studies on electron-doped iron-oxypnictides La1111 compounds that the increase in $T_c$ is not due to the development of AFM spin fluctuations, but because the structural parameters, such as the bond angle $\alpha$ of the FeAs$_4$ tetrahedron and the a-axis length, approach each optimum value. Based on these results, we propose that a stronger correlation in HTSC than in FeAs-based superconductors may make $T_c$ higher significantly. 

\end{abstract}


\end{frontmatter}



\section{Introduction}
\label{}
Despite more than 24 years of intensive research, the origin of high-temperature copper-oxide superconductivity (HTSC) has not yet been well understood. HTSC can be observed on a CuO$_2$ plane when an antiferromagnetic Mott insulator is doped with mobile carriers. A strong relationship between antiferromagnetism (AFM) and superconductivity (SC) is believed to be the key to understanding the origin of remarkably high SC transition temperatures~\cite{Chen,Giamarchi,Inaba,Anderson1,Lee1,Zhang,Himeda,Kotliar,Paramekanti1,Lee2,Demler,Yamase,Paramekanti2,Shih1,Shih2,Senechal,Capone,Ogata,Pathak}. As a matter of fact, site-selective $^{63}$Cu-NMR studies on multilayered cuprates revealed that AFM uniformly coexists with SC in a single CuO$_2$ plane~\cite{Mukuda2006,Shimizu2007,Mukuda2008,Shimizu2009PRB,Shimizu2009JPSJ,Mukuda2010}; Square-type inner CuO$_2$ planes (IPs) are located so far from the charge reservoir layers (CRLs), in which the disorder is introduced by the chemical substitution, exhibiting homogeneous hole doping. In the previous papers, the  NMR investigations of five-layered ($n$=5) compounds~\cite{Mukuda2006,Mukuda2008,Mukuda2010} and  four-layered ($n$=4) compounds~\cite{Shimizu2007,Shimizu2009PRB,Shimizu2009JPSJ}
have revealed that the AFM quantum critical point (QCP) takes place around the hole densities $p_{c}\sim$ 0.17 and $\sim$ 0.15, respectively. These $p_{c}$s are significantly larger than $p_{c}\sim$~0.02 for La$_{2-x}$Sr$_x$CuO$_4$ ($n$=1:LSCO)~\cite{LSCO} and $p_{c}\sim$~0.055 for YBa$_2$Cu$_3$O$_{6+y}$($n$=2:YBCO)~\cite{YBCO}. Furthermore, $p_{max}\sim$~0.22 giving rise to the maximum of $T_{\rm c}$ for the $n=4$ and $n=5$ compounds is also larger than $p_{max}\sim$ 0.16 for the $n$=2 compounds. 

Upon doping with holes or~electrons$-$e.g., by replacing some of the out-of-plane atoms with different oxidation states$-$ AFM is destroyed and the materials become superconductors at lower temperatures. For example, in LSCO, La$^{3+}$ is exchanged with Sr$^{2+}$, which increases the planar CuO$_2$ hole density.  The situation with the in-plane hole density and distribution in other superconducting cuprates is even less clear. Indirect chemical methods like solid solutions~\cite{1}, semiempirical bond valence sums determined from structural bond lengths~\cite{2,3,4} or methods based on the Fermi surface topology~\cite{5} are used to determine the hole density. To the best of our knowledge there are no direct physical measurements that can establish a phase diagram in terms of the planar CuO$_2$ hole density. 
%
\begin{figure*}[tbp]
\centering
\includegraphics[width=14.3cm]{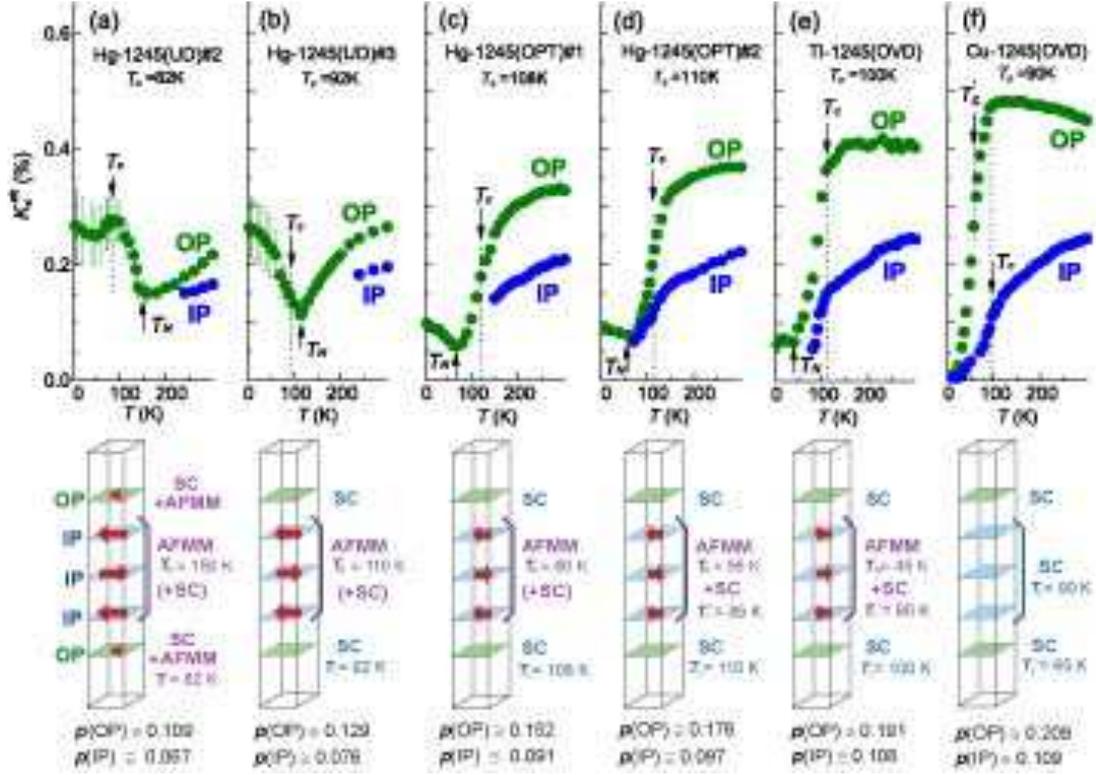}
\caption[]{\footnotesize (Color online) 
Temperature dependence of $K_s^{ab}$ for (a):Hg-1245(UD)\#2 ($T_c$=82K)~\cite{Mukuda2010}, (b):Hg-1245(UD)\#3 ($T_c$=92K)~\cite{Mukuda2010}, (c):Hg-1245(OPT)\#1 ($T_c$=108K)~\cite{Kotegawa2004}, (d):Hg-1245(OPT)\#2 ($T_c$=110K)~\cite{Mukuda2008}, (e):Tl-1245(OVD) ($T_c$=100K)~\cite{Mukuda2008,Kotegawa2004}, and (f):Cu-1245(OVD) ($T_c$=90K)~\cite{Kotegawa2001}. The hole density $p$ for IPs and OPs is independently estimated using the renewed relationship of $p$ versus $K_s$(RT), $p=$0.502$K_s$(RT)$-$0.014~\cite{Shimizu2010JPSJ} (see the text). Layer-dependent physical properties are summarized in lower panels.
}
\label{Knightshift}
\end{figure*}
%

Under this situation, using NMR/NQR (nuclear quadrupole resonance) that can distinguish between the various atoms in the unit cell, Zheng {\it et al.}  had attempted to interpret the nuclear quadrupole interaction for planar Cu and O, in terms of the local hole densities (see, e.g., Ref.\cite{NQR1} and references cited in it). This is indeed a sensible approach since the electric field gradient (EFG) at a nuclear site is very sensitive to changes in the local charge distribution, namely the planar Cu and O EFG's show a pronounced doping dependence. In these analyses, Zheng {\it et al.}  used the experimental data available until 1995 with the help of atomic and cluster calculations to connect the hole densities to the EFG tensors for Cu and O. Furthermore, in order to estimate each hole density at IPs and pyramidal outer planes (OPs) in multilayered compounds, we have used an empirical relationship between the spin component in  Knight shift $K_{s}$(RT) at room temperature and the planar CuO$_2$  hole density ($p$) determined with NMR/NQR. 

Meanwhile, Hasse {\it et al.}~\cite{NQR2} performed an almost model independent analysis to relate the hole densities to the EFG tensors without relying on calculations and with the extensive experimental data until 2004 including those for the parent compounds of LSCO and YBCO. On the basis of this more precise analysis to relate the hole densities to the EFG tensors than before~\cite{NQR1}, Shimizu {\it et al.}~\cite{Shimizu2010JPSJ}, have constructed a phase diagram of the $n$=2 apical-fluorine compounds Ba$_2$CaCu$_2$O$_4$(F,O)$_2$(0212F) using a renewed empirical relationship of $K_{s}$(RT) versus $p$. Remarkably, thus obtained phase diagram has coincided with the well established ones for the $n$=2 YBCO and Bi$_2$Ba$_2$CaCu$_2$O$_{8+y}$~(Bi2212), exhibiting the maximum of $T_{\rm c}$ at $p_{max}$$\sim$~0.16. Thus, the universal phase diagram has been established on the $n$=2 compounds independent of the methods to estimate the hole densities. Motivated by this fact, in the first part of the paper, we reexamine the phase diagrams of AFM and SC in multilayered compounds reported extensively in the literatures~\cite{Mukuda2006,Shimizu2007,Mukuda2008,Shimizu2009PRB,Shimizu2009JPSJ} 

In the second part, we review extensive NMR studies on electron-doped iron-oxypnictides La1111 compounds in order to shed light on novel normal-state properties and their relevance with unique SC characteristics and address remarkable differences from those in HTSC. 

\section{High-$T_c$ cuprates}
\subsection{NMR results on five-layered ($n$=5) compounds}
Fig.~\ref{Knightshift} shows temperature ($T$) dependence of the spin component of Knight shift $K^{ab}_s(T)$~($\perp$ the c-axis) for $n$=5 compounds such as (a) Hg-1245(UD)\#2 ($T_c$=82K) \cite{Mukuda2010}, (b) Hg-1245(UD)\#3 ($T_c$=92K) \cite{Mukuda2010}, (c) Hg-1245(OPT)\#1 ($T_c$=108K) \cite{Kotegawa2004}, (d) Hg-1245(OPT)\#2 ($T_c$=110K) \cite{Mukuda2008}, (e) Tl-1245(OVD) ($T_c$=100K) \cite{Mukuda2008,Kotegawa2004}, and (f) Cu-1245(OVD) ($T_c$=90K) \cite{Kotegawa2001}. 
$K_s$(RT) at room temperature is proportional to the planar CuO$_2$ hole density $p$~\cite{NQR1,Kotegawa2001}. Here, $p$ for IPs and OPs is independently estimated using the renewed relationship of $p$ versus $K_s$(RT), $p$=0.502$K_s$(RT)$-$0.014~\cite{Shimizu2010JPSJ}.

In the $T$ dependence of $K^{ab}_s(T)$(OP), an unusual upturn was observed below temperatures marked by up-arrows in Figs.1(a)-(e), below which the linewidth of $^{63}$Cu-NMR spectra at OPs steeply increases.  
Note that this upturn in $K^{ab}_s(T)$ at OPs was associated with an onset of AFM order at IPs, which was also corroborated by a critical enhancement of $1/T_1$ in the previous studies on the other Hg-1245 compounds~\cite{Mukuda2008,Kotegawa2004}. 
As a result of reduction of hole densities, it was demonstrated that the N\'eel temperature $T_N$ at IPs increases from $T_N$=45 K for (e)~Tl-1245(OVD) ($T_c$=100 K)~\cite{Mukuda2008,Kotegawa2004} to $T_N$=150 K for (a)~Hg-1245(UD)\#2 ($T_c$=82 K)~\cite{Mukuda2010}. Layer-dependent physical properties are summarized in lower panels of Figs~\ref{Knightshift}(a-f). 

Spontaneous AFM moments $M_{\rm AFM}$(IP) at IPs and OPs for the $n$=5 compounds were determined with the zero-field Cu-NMR at 1.5 K. This is because  an internal field ($H_{\rm int}$) is induced by $M_{\rm AFM}$(IP) at the Cu sites in IPs and OPs. $M_{\rm AFM}$(IP) is evaluated  by using the relation $H_{\rm int}$(IP)=$|A_{\rm hf}$(IP)$|M_{\rm AFM}$(IP) with the hyperfine coupling constant $A_{\rm hf}$(IP)=$-$20.7 T$/\mu_{\rm B}$~\cite{Kotegawa2004}. The details are referred to the literatures~\cite{Mukuda2008,Mukuda2010,Kotegawa2004}. The values of $M_{\rm AFM}$(IP) ranging in 0.04$-$0.20$\mu_{\rm B}$ are significantly reduced by the mobile holes from 0.5$-$0.7$\mu_{\rm B}$ in undoped cuprates~\cite{Vaknin1987,Vaknin1989}. This is consistent with the conclusion of neutron diffraction measurement on (a)~Hg-1245(UD)\#2 ($T_c$=82~K)~\cite{Mukuda2010}; their magnetic moments were suggested to be less than 0.3$\mu_B$ if any, since a Bragg intensity, that enables to estimate $M_{\rm AFM}$, was not apparent~\cite{CHLee}.  The NMR outcome on (a)~Hg-1245(UD)\#2 ($T_c$=82~K) ensures that the SC with $T_c$=82 K uniformly coexists with the AFM order of $M_{\rm AFM}$(OP)$\sim$~0.05$\mu_{\rm B}$ at OPs~\cite{Mukuda2010}.  This result convinces along with the previously reported ones that the uniform coexistence of AFM and SC is a general event for homogeneously underdoped CuO$_2$ plane. Coexisting state of AFM and SC order are known in multi-band materials where the AFM order occurs in localized $d$- and $f$-bands and the SC order in $s$- and $p$-bands. However, HTSC is characterized by only a single electron fluid composing of~Cu-O hybridized electrons. In such case, one may expect competition rather than coexistence between the two types of orders, but it is worth noting that the $T_c$ of the coexisting phase is relatively high ($T_c\sim$~82 K), revealing the close relationship between AFM order and SC order.
%
\begin{figure}[ht]
\centering
\includegraphics[width=9cm]{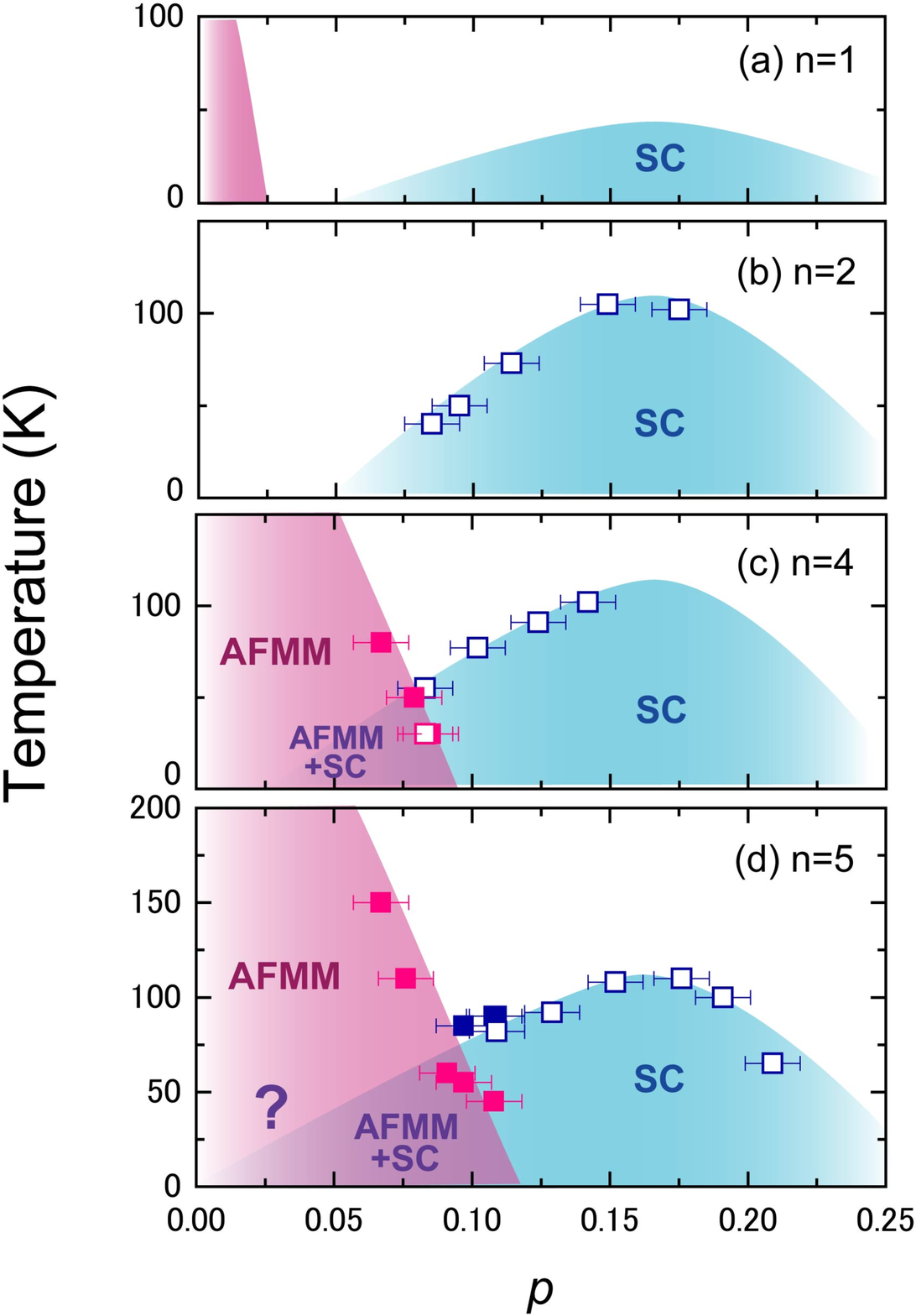}
\caption[]{\footnotesize (Color online) The phase diagrams of AFM and SC for (a)~the $n$=1 LSCO, (b)~the $n$=2 and (c)~the $n$=4 apical fluorine systems~\cite{Shimizu2007,Shimizu2009PRB,Shimizu2009JPSJ,Shimizu3}, and (d)~the $n$=5 compounds~\cite{Mukuda2008,Mukuda2010}.  In these phase diagrams, $T_{\rm N}$ and $T_{\rm c}$ are plotted as the function of $p$. Here the solid and empty squares correspond to the respective data for IPs and OPs.}
\label{fig:PhaseDiagram}
\end{figure}

\subsection{Phase diagram of CuO$_2$ plane}
In Figs.~\ref{fig:PhaseDiagram}(a)-(d), we present the phase diagrams of AFM and SC for (a)~the $n$=1 LSCO, (b)~the $n$=2~\cite{Shimizu2010JPSJ} and (c)~the $n$=4 apical fluorine systems~\cite{Shimizu2007,Shimizu2009PRB,Shimizu2009JPSJ,Shimizu3}, and (d)~the $n$=5 compounds~\cite{Mukuda2008,Mukuda2010}.  In these phase diagrams, $T_{\rm N}$ and $T_{\rm c}$ are plotted as the function of $p$. Here the solid and empty squares correspond to the respective data for IPs and OPs. It should be noted that the phase diagram of the $n$=2 apical fluorine compounds does not reveal an AFM order~\cite{Shimizu2010JPSJ}. This phase diagram resembles the well-established phase diagram of YBCO~\cite{YBCO} in which  AFM order totally collapses at $p_{c}\sim$~0.055. However, an AFM order takes place for the $n$=4 and 5 compounds~\cite{Shimizu2007,Mukuda2008,Shimizu2009PRB,Shimizu2009JPSJ,Mukuda2010,Shimizu3}. It is noteworthy that $p_{c}$ at which AFM order collapses increases from $p_{c}$=0.10 to 0.12 as $n$ increases from 4 to 5.
This result suggests that interlayers magnetic coupling $\sqrt{J_cJ_{\rm out}(n)}$ becomes stronger with increasing $n$, which stabilizes the AFM order. Here, $J_{\rm c}$ is magnetic coupling between interunit cells, which is independent of $n$, but $J_{\rm out}(n)$ is magnetic coupling between intraunit cells, which increases with $n$. 
The characteristic features are summarized as; (1)~the respective $p_{c}\sim$~0.10 and 0.12 for the $n$=4 and 5 compounds are significantly smaller than $p_{c}\sim$~0.15 and 0.17 which were estimated in the previous papers~\cite{Shimizu2007,Mukuda2008,Shimizu2009PRB,Shimizu2009JPSJ,Mukuda2010,Shimizu3}. This is because the previous analysis had overestimated $p$ using the experimental data available data until 1995 with the help of atomic and cluster calculations to connect the hole densities to the EFG tensors for Cu and O. (2)~The maximum of $T_{\rm c}$ for each compound takes place universally around $p_{max}\sim$~0.16. It is noteworthy that the highest $T_c$ happens for the $n$=3 compound as expected.    

\subsection{The $T$=0 phase diagram of AFM and superexchange interaction with doping}
\begin{figure}[ht]
\centering
\includegraphics[width=9cm]{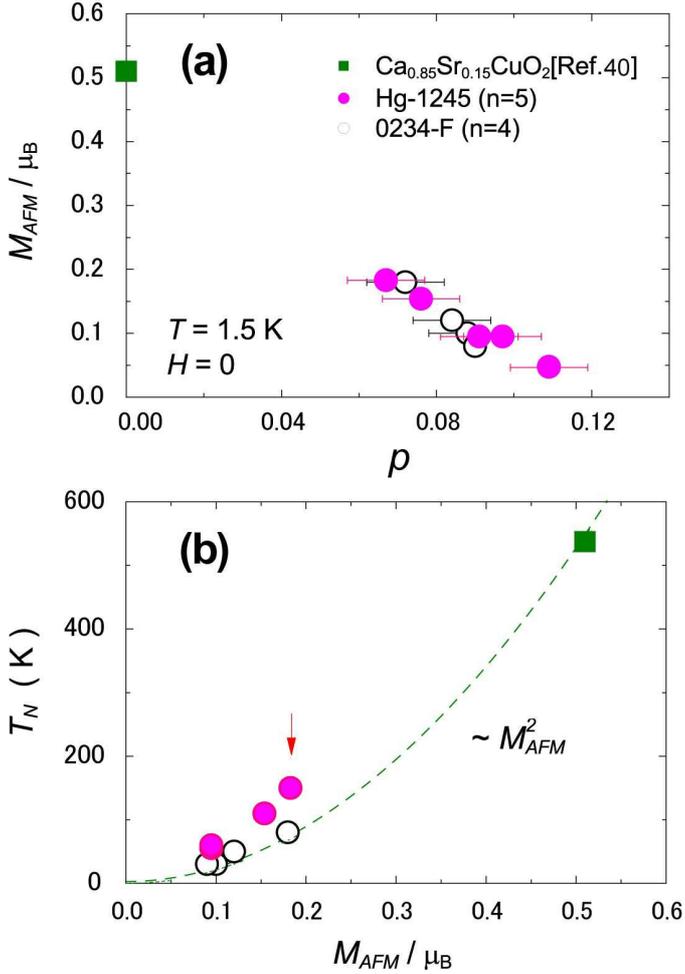}
\caption[]{\footnotesize (Color online) (a)~Plot of $M_{\rm AFM}$ at $T=$1.5 K versus $p$~\cite{Mukuda2008,Shimizu2009JPSJ,Vaknin1989}. $p_c$ at $M_{\rm AFM}$=0 is around $\sim$ 0.11. (b) Plot of $T_N$ versus $M_{\rm AFM}$~\cite{Shimizu2007,Shimizu2009PRB,Shimizu2009JPSJ,Shimizu3,Mukuda2008,Mukuda2010,Vaknin1989}. The broken curve shows $T_N\propto M_{\rm AFM}^2$ with $T_N$=537~K at $M_{\rm AFM}$=0.51$\mu_B$ in Ca$_{0.85}$Sr$_{0.15}$CuO$_{2}$~\cite{Vaknin1989} where $n=\infty$. The arrow points to $M_{\rm AFM}\sim$~0.18$\mu_B$.}
\label{Mvsp}
\end{figure}

Fig.~\ref{Mvsp}(a) shows a plot of the $M_{\rm AFM}$ at $T$=1.5~K versus $p$~\cite{Shimizu2007,Mukuda2008,Shimizu2009PRB,Shimizu2009JPSJ,Mukuda2010,Shimizu3}, where the datum at $p$=0 for a Mott insulator is cited from the infinite-layered ($n=$ $\infty$) AFM-Mott insulator Ca$_{0.85}$Sr$_{0.15}$CuO$_{2}$ with $T_N$=537~K and $M_{\rm AFM}$=0.51$\mu_B$~\cite{Vaknin1989}. The phase diagram of $M_{\rm AFM}$ versus $p$ presented here is totally consistent with the $T$=0 phase diagrams in a single CuO$_2$ plane which were theoretically addressed thus far in terms of either the t-J model~\cite{Chen,Giamarchi,Lee1,Himeda,Kotliar,Paramekanti1,Paramekanti2,Shih1,Shih2,Ogata,Pathak}, or the Hubburd model in a strong correlation regime~\cite{Senechal,Capone}. A critical hole density $p_c$ at $M_{\rm AFM}$=0 taking place in the range 0.10$<p_c<$0.12 is in quantitative agreement with the calculations.
Fig.~\ref{Mvsp}(b), which shows a plot of $T_N$ versus $M_{\rm AFM}$, helps us to gain further insight into the $p$ dependence of in-plane exchange interaction $J_{\rm in}(p)$.  In this figure, data are presented with respect to Ca$_{0.85}$Sr$_{0.15}$CuO$_{2}$ having $n=\infty$ for $p$=0~\cite{Vaknin1989}, the $n$=4 compound~\cite{Shimizu2009JPSJ}, and the $n$=5 compound~\cite{Mukuda2008,Mukuda2010}. On the basis of the mean-field approximation of localized spins, we assume that $T_N\propto$~$M_{\rm AFM}^2$, and $J_{\rm out}(\infty)$ and $J_{\rm in}(\infty)$ for Ca$_{0.85}$Sr$_{0.15}$CuO$_{2}$ stay constant regardless of $p$. Using $T_N$=537~K at $M_{\rm AFM}$=0.51~$\mu_B$ for Ca$_{0.85}$Sr$_{0.15}$CuO$_{2}$, a graph of $T_N\propto$~$M_{\rm AFM}^2$ is plotted, which is indicated by the broken curve in Fig.~\ref{Mvsp}(b). First, we notice that even though $M_{\rm AFM}\sim$~0.18~$\mu_B$ remains constant (shown by an arrow in Fig.~\ref{Mvsp}(b)), $T_N(n)$ increases due to the increase in $J_{\rm out}(n)$ as $n$ increases from 4 to 5, namely, $J_{\rm out}(4)<J_{\rm out}(5)$. Although $J_c$ and $J_{\rm out}(n)$ values for $n$=4 and 5 are always smaller than $J_{\rm out}(\infty)$ in Ca$_{0.85}$Sr$_{0.15}$CuO$_{2}$, $T_N(n)$$\sim$$M_{\rm AFM}^2(p)[J_{\rm in}(p)\sqrt{J_cJ_{\rm out}(n)}]^{1/2}$ are larger than $T_N(\infty)$$\sim$$M_{\rm AFM}^2(p)[J_{\rm in}(\infty)J_{\rm out}(\infty)]^{1/2}$ with $T_N$=537 K at $M_{\rm AFM}$=0.51$\mu_B$ and $p$=0 (see the broken curve in Fig.~\ref{Mvsp}(b));~we also obtain an unexpected relation, i.e., $J_{\rm in}(p)>J_{\rm in}(\infty)\sim$~1300~K. The two experimental~relationships$-$the plot of $M_{\rm AFM}$ versus $p$ shown in Fig.~\ref{Mvsp}(a) and the plot of $T_N$ versus $M_{\rm AFM}$ shown in Fig.~\ref{Mvsp}(b)$-$ suggest that the AFM ground state in the homogeneously hole-doped CuO$_2$ layers is determined by $p$, $\sqrt{J_cJ_{\rm out}(n)}$ and $J_{\rm in}(n,p)$, which is larger than $J_{\rm in}(\infty)\sim$~1300~K. It is surprising that $J_{\rm in}(p)$ becomes larger for hole-doped AFM CuO$_2$ planes than for undoped AFM-Mott insulators.  Mean-field theories of HTSC used to consider the Heisenberg superexchange interaction $J_{\rm in}$ as the source of an instantaneous attraction that led to pairing in a d-wave state~\cite{Anderson2}. The present outcomes may support such a scenario experimentally as far as the underdoped region is concerned, where AFM and SC uniformly coexist in a CuO$_2$ plane.

\subsection{Summary on high-$T_c$ cuprates}
\label{}
The site-selective NMR studies on multilayered compounds has led us to the following remarkable conclusions based on the more precise evaluation of local hole density ($p$)~\cite{NQR2} than before~\cite{NQR1}:
\begin{enumerate}
\item The $T$=0 phase diagram of AFM which is experimentally determined in Fig.~\ref{Mvsp}(a) is in quantitative agreement  with the $T$=0 phase diagrams in terms of either the~t-J model~\cite{Chen,Giamarchi,Inaba,Anderson1,Lee1,Zhang,Himeda,Kotliar,Paramekanti1,Lee2,Demler,Yamase,Paramekanti2,Shih1,Shih2,Ogata,Pathak}, or the Hubburd model in the strong correlation regime~\cite{Senechal,Capone}.
\item  $T_N$ increases as the intraunit cell magnetic coupling $J_{\rm out}(n)$ increases with increasing $n$.
\item The in-plane exchange interaction $J_{\rm in}(p)$ for the $n=$ 4 and 5 compounds is larger than $J_{\rm in}(\infty)\sim$~1300 K for the infinite-layered AFM-Mott insulator Ca$_{0.85}$Sr$_{0.15}$CuO$_{2}$.
\end{enumerate}
While noting that $T_c$ exhibits the maximum at $p_{max}\sim$~0.16 close to $p_c\sim$~0.11 at the AFM-QCP regardless of $n$, the results presented here demonstrate that both AFM and SC are mediated by AFM interaction, which also acts the glue for the Cooper pairs~\cite{Anderson2} and hence makes $T_c$ of HTSC very high. It is highlighted that the on-site Coulomb repulsive interaction is almost unchanged with doping, being almost the same as in AFM-Mott insulators. Further theoretical study is desired to address whether $J_{in}$($p$) becomes larger than for AFM-Mott insulators even though mobile holes are doped. 

\section{NMR studies on electron-doped iron-oxypnictides}

A new class of SC compounds including iron element has been discovered~\cite{Kamihara}. In all the cases of this class, Fe-3d conduction electrons are likely to form Cooper pairs and responsible for SC. However, the mechanism of SC is not well understood and is under extensive debates. In the family with ZrCuSiAs-type structure (called 1111 hereafter), SmFeAs(O,F) has shown the highest SC
critical temperature $T_c$ =56K~\cite{Ren} when fluorine is substituted for 20\% of oxygen as electron doping, while BaFe$_2$As$_2$ with ThCr$_2$Si$_2$-type structure (called 122) has indicated the highest $T_c$=38 K, when potassium is substituted for 40\% of Ba as hole doping~\cite{Rotter}. It is not yet addressed why $T_c$ is highest in the Ln1111 system. 
Recently, it has been reported that $T_c$ can be increased by either a Y or a H substitution in the La1111 system without replacing magnetic rare-earth elements~\cite{Shirage,Tropeano,MiyazawaH}, in which the angle $\alpha$ of~As-Fe-As bonding and the $a$-axis length approach those of the Nd1111 system with the highest $T_c$ to date~\cite{C.H.Lee,Shirage,Ren2}. 
There is a way to address the intimate correlation between the evolution of electronic state and the local structure of the FeAs$_4$ tetrahedron. 

Here, we review on normal-state and SC characteristics on La1111 compounds through extensive NMR measurements~\cite{MukudaLa1111Y}. We focus on evolutions of the electronic state derived by Y- and H-substitution in the La1111 compounds and address why $T_c$s in these compounds are enhanced. The substitutions of yttrium (Y) and hydrogen (H) into~optimally-doped LaFeAsO$_{1-y}$ (La1111(OPT)) increase $T_c$=28 K to $T_c$=34 K for La$_{0.8}$Y$_{0.2}$FeAsO$_{1-y}$ (La$_{0.8}$Y$_{0.2}$1111) and $T_c$=32 K LaFeAsO$_{1-y}$H$_{x}$(La1111(H)), respectively~\cite{Shirage,Tropeano,MiyazawaH}. 

In the normal state, the measurements of nuclear spin-relaxation rate $1/T_1$~\cite{MukudaLa1111Y} revealed that the electron doping level in La$_{0.8}$Y$_{0.2}$1111 is close to that of La1111(OPT) where $1/T_1T$ decreases upon cooling without the development of AFM spin fluctuations~\cite{Terasaki,MukudaNQR,MukudaFe2}, confirming that the Y$^{3+}$ substitution for La$^{3+}$ does not change the doping level. As a result, the reason that $T_c$ increases in La$_{0.8}$Y$_{0.2}$1111 is not due to the change in its doping level, but because the structural parameters, such as the bond angle $\alpha$ of the FeAs$_4$ tetrahedron and the $a$-axis length, approach each optimum value. 
By contrast, in  La1111(H), the $(1/T_{1}T)$ is rather comparable with that for the underdoped LaFeAsO$_{0.93}$F$_{0.07}$ ($T_c=$ 22.5 K) where $1/T_1T$ stays constantly upon cooling without the development of AFM spin fluctuations as well~\cite{Nakai2}.  These results imply that the La1111(H) is in an underdoped regime.

\begin{figure}[htbp]
\begin{center}
\includegraphics[width=8cm]{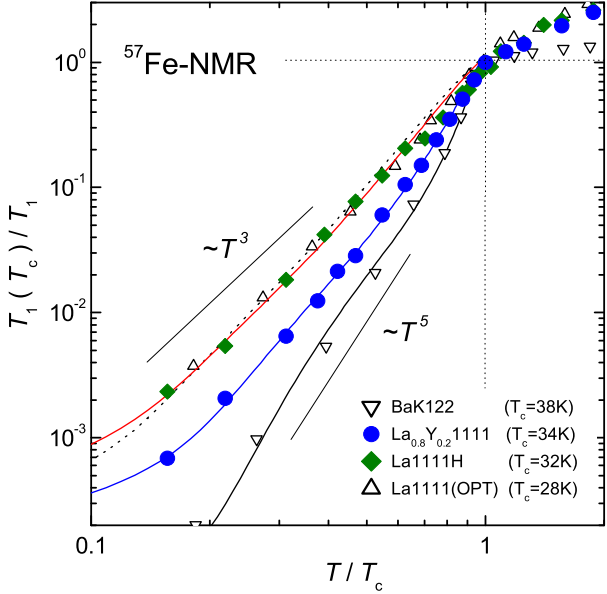}
\end{center}
\caption[]{\footnotesize (color online)
$T$ dependence of $^{57}$Fe-$(T_1(T_c)/T_1)$ normalized  at $T_c$ in the SC state of La$_{0.8}$Y$_{0.2}$1111 and La1111(H), along  with the results reported in the literature on BaK122 and La1111(OPT) \cite{Yashima}. 
The multiple fully-gapped $s_\pm$-wave model allows us to deduce the parameters for all compounds listed up in Table \ref{table1} through reproducing the $T$ dependence of $^{57}$Fe-$(T_1(T_c)/T_1)$. 
}
\label{fig:FeNMRT1SC}
\end{figure}

Fig.~\ref{fig:FeNMRT1SC} indicates the $\sl{T}$ dependence of $^{57}(T_1(T_c)/\sl{T}_{1})$ normalized at their $T_c$s.   The $^{57}(T_{1}(T_c)/\sl{T}_{1})$ in La$_{0.8}$Y$_{0.2}$1111 ($T_c$=34 K) decreases as nearly $\sim T^4$ upon cooling below $\sl{T}_{c}$, which differs from either the $\sl{T}^{3}$ in La1111(OPT)~\cite{MukudaNQR,Terasaki} or the $\sl{T}^{5}$ in Ba$_{0.6}$K$_{0.4}$Fe$_{2}$As$_{2}$(BaK122)~\cite{Yashima}. It is reinforced that a common power-law like $T$ dependence of $1/T_1$ is not evident at all among Fe-based new superconductors. These results contrast with the behavior of $1/T_1\sim T^3$ observed commonly in high-$T_c$ cuprates which are the $d$-wave superconductor with line-node gap. 

Meanwhile, it was shown in the literature~\cite{Yashima} that the relaxation behaviors of both the $\sl{T}^{3}$ in La1111(OPT)~\cite{Terasaki,Nagai} and the $\sl{T}^{5}$ in Ba$_{0.6}$K$_{0.4}$Fe$_{2}$As$_{2}$(BaK122)~\cite{Yashima} are consistently reproduced in terms of the multiple fully-gapped $s_\pm$-wave model. This model is also applicable for understanding the SC characteristics of La$_{0.8}$Y$_{0.2}$1111 and La1111(H) as follows. In model A, respective Fermi surfaces (FS1 and FS2) have an isotropic gap and an anisotropic gap denoted as $\Delta^{FS1}=\Delta_0$ and $\Delta^{FS2}(\phi)=(\Delta_0+\Delta_{min})/2 + (\Delta_0-\Delta_{min})\cos 2\phi/2$~\cite{Nagai}. In model B, both Fermi surfaces have the isotropic gaps as $\Delta^{FS1}= \Delta_L$ and $\Delta^{FS2}=\Delta_S$.  According to the model A(B) in the literature~\cite{Nagai,Yashima}, a fraction of the density of states (DOS) at FS1 is taken as $N_{FS1}/(N_{FS1}+N_{FS2})$=0.4 (0.7) and $\Delta_{min}$ = 0.25 $\Delta_0$ ($\Delta_{S}/\Delta_L$=0.35).  Furthermore, the coherence factor is neglected on the assumption that the interband scattering between the sign reversal gaps becomes dominant for the relaxation process. In fact, the $T$ dependence of $^{57}(T_1(T_c)/\sl{T}_{1})$ of La$_{0.8}$Y$_{0.2}$1111 is consistently reproduced using model B with the parameters $2\Delta_L/k_BT_c$ = 6.9$(\Delta_{S}/\Delta_L$=0.35) and a smearing factor $\eta=0.04\Delta_L$, as is shown by the solid line in Fig.~\ref{fig:FeNMRT1SC}. Here, the smearing factor ($\eta/\Delta_L$) reflects a damping effect of quasiparticles due to impurity scattering. Notably, $2\Delta_L/k_BT_c$ = 6.9 in La$_{0.8}$Y$_{0.2}$1111 is larger than the 4.4 in La1111(OPT), revealing that the Y substitution into La1111 increases $T_c$ up to 34 K in association with a strong-coupling effect to mediate the Cooper pairs. Despite the fact that Y substitution introduces some disorder as suggested by $\eta$ (see Table~\ref{table1}) in La$_{0.8}$Y$_{0.2}$1111 being larger than in La1111(OPT), it is noteworthy that the strong-coupling effect enhances $T_c$. 
%
\begin{table}[htbp]
\centering
\caption[]{Evolutions of SC gap ($\Delta_L$) and smearing factor ($\eta$) obtained from the analyses of $^{57}$Fe-$(T_1(T_c)/T_1)$ for La$_{0.8}$Y$_{0.2}$1111 and La1111(H) assuming the multiple fully-gapped $s_\pm$ wave model, which was applied to BK122 and La1111(OPT)~\cite{Yashima}(see text). Here $\Delta_L$ represents the larger gap in model B assuming two full gaps. The $\eta$ reflects a damping effect of quasiparticles due to impurity scattering, so that the $\eta/\eta_0$ represents the disorder effect introduced by Y or H substitutions, where $\eta_0$ is defined as that of La1111(OPT). The angle $\alpha$ of As-Fe-As bonding is evaluated from the $a$-axis length at room $T$ and a fixed bond length of Fe-As $\sim$ 2.41\AA~\cite{C.H.Lee}.}
\begin{tabular}{lccccc}
\hline
           & $T_c$ &         $a$(\AA)         & $2\Delta_L/k_BT_c$  & $\eta/\eta_0$ \\
           & (K)   &  ($\alpha$($^\circ$))    & ($\eta/\Delta_L$)   &         \\
\hline
BaK122\cite{Rotter,Yashima} & 38 &  3.917    &  9.4                 & 0.64 \\
                            &    &   (109.7) &  (0.015)             &      \\
La$_{0.8}$Y$_{0.2}$1111     & 34 &  4.004    &  6.9                 & 1.3  \\ 
                            &    &   (112.3) &  (0.04)              &      \\
La1111(H)                   & 32 & 3.989     &  $\sim$4.7           & $\sim$1.3  \\    
                            &    &   (111.7) &  ($\sim$0.06)        &      \\
La1111(OPT)\cite{Yashima}   & 28 & 4.023     &  4.4                 & 1    \\
                            &    &  (113.2)  &  (0.5)               &      \\
\hline
\end{tabular}
\label{table1}
\end{table} 

As for La1111H, it is possible that $^{57}(T_1(T_c)/\sl{T}_{1})$ is also reproduced roughly in terms of model B with the parameters $2\Delta_L/k_BT_c\sim 4.7$ and $\eta\sim 0.06\Delta_L$ similar to those in La1111(OPT), as shown by the solid line in Fig.~\ref{fig:FeNMRT1SC}. 
Despite the fact that the electron carrier density is lower than in La1111(OPT) and the disorder is significantly introduced into the Fe site, which is deduced from $\eta/\eta_0\sim1.3$, the $T_c$ is enhanced up to 32 K. As a result, the reason that $T_c$ increases in La1111(H) may be closely related to the slight increase of the SC gap as well.

In the SC state, the measurements of $1/T_1$ have revealed in terms of a multiple~fully-gapped $s_\pm$-wave model that the SC gap and $T_c$ in La0.8Y0.21111 becomes larger than those in La1111(OPT) without any change of doping level, and the $T_c$ in La1111H increases even though the carrier density is decreased and some disorder is significantly introduced. As a consequence, it is highlighted that the reason that $T_c$ is increased in both compounds is due to neither the change in doping level nor the development of AFM spin fluctuations, but because the structural parameters approach each optimum value to increase $T_c$  for the bond angle $\alpha$ of the FeAs$_4$ tetrahedron and the a-axis length. Systematic spectroscopies of SC property on Ln(1111) systems with $T_c > 50$ K  are highly desired in the future in order to unravel the Fermi surface topologies and their relevance with SC gap structures.

\section{Concluding remarks}
The underdoped high-$T_c$ cuprates revealing the uniform coexistence of AFM and SC are in the strong correlation regime which is unchanged even with mobile holes. This~{\it Mott physics} plays a vital role for mediating the Cooper pairs to make $T_c$ of HTSC very high. On the other hand, in iron-oxypnictides, the role of AFM interaction in $T_c$ is even less clear and the correlation effect is not clearly visible experimentally~\cite{spectroscopy}. Rather, a large $d-p$ hybridization in the 1111 family generates a large {\it band-insulating}-like pseudogap ({\it hybridization gap}).
In this context, we remark that the stronger correlation in HTSC than in FeAs-based superconductors may make $T_c$ higher significantly.

This work was supported by a Grant-in-Aid for Specially Promoted Research (20001004) and by the Global COE Program (MEXT).

\end{document}